\documentclass{jpp}

\usepackage{graphicx}
\usepackage{natbib}
\usepackage[colorlinks,urlcolor=blue,citecolor=blue,linkcolor=blue]{hyperref}
\usepackage{amsmath}
\usepackage{subcaption}
\usepackage{gensymb}

\ifCUPmtlplainloaded \else
  \checkfont{eurm10}
  \iffontfound
    \IfFileExists{upmath.sty}
      {\typeout{^^JFound AMS Euler Roman fonts on the system,
                   using the 'upmath' package.^^J}%
       \usepackage{upmath}}
      {\typeout{^^JFound AMS Euler Roman fonts on the system, but you
                   dont seem to have the}%
       \typeout{'upmath' package installed. JPP.cls can take advantage
                 of these fonts, if you use 'upmath' package.^^J}%
      }
  \else
  \fi
\fi

\ifCUPmtlplainloaded \else
  \checkfont{msam10}
  \iffontfound
    \IfFileExists{amssymb.sty}
      {\typeout{^^JFound AMS Symbol fonts on the system, using the
                'amssymb' package.^^J}
       \usepackage{amssymb}%
         \let\leq=\leqslant
         \let\geq=\geqslant
      }{}
  \fi
\fi

\ifCUPmtlplainloaded \else
  \IfFileExists{amsbsy.sty}
    {\typeout{^^JFound the 'amsbsy' package on the system, using it.^^J}%
     \usepackage{amsbsy}}
    {\providecommand\boldsymbol[1]{\mbox{\boldmath $##1$}}}
\fi

\usepackage{graphicx}
\usepackage{wrapfig}
\usepackage{natbib}
\usepackage{url}
\usepackage{enumitem}

\usepackage{graphics}

\newcommand{\be}{\begin{equation}}
\newcommand{\ee}{\end{equation}}
\newcommand{\nn}{\mbox{} \nonumber \\ \mbox{} }
\newcommand{\ba}{\begin{eqnarray}}
\newcommand{\ea}{\end{eqnarray}}
\newcommand{\om}{\omega}
\newcommand{\Alfven}{Alfv\'{e}n }

\newcommand{\E}{{\bf E}}
\newcommand{\B}{{\bf B}}
\newcommand{\J}{{\bf J}}

\newcommand{\Bf}{{magnetic field}}

\newcommand{\Ef}{{electric  field}}

\newcommand{\NS}{neutron star}
\newcommand{\NSs}{{neutron stars}}
\newcommand{\EM}{electromagnetic}

\newcommand{\ms}{magnetosphere}
\newcommand{\mss}{magnetospheres}

\begin{document}

\title{Relativistically-strong  electromagnetic waves in magnetized plasmas}
\author{Maxim Lyutikov\\
Department of Physics and Astronomy,  Purdue University, \\
 525 Northwestern Avenue,
West Lafayette, IN
47907-2036 }

\maketitle

\begin{abstract}
 Using a two-fluid approach, we consider the properties of relativistically nonlinear (arbitrary $a_0$), circularly polarized  \EM\ waves propagating along \Bf\  in electron-ion and pair plasmas. Dispersion relations depend on how wave intensity scales with frequency, \eg $a_0 (\om)$.  For superluminal branches, the nonlinear effects reduce the cut-off frequency, while the general form of the dispersion relations $\om(k)$ remains similar to the linear case. For subluminal waves, whistlers and  Alfven, a new effect appears: dispersion curves effectively terminate at finite  $\om^\ast  - k^\ast$, { where  the group velocity becomes zero.}   Qualitatively,  subluminal modes with fluctuating \Ef\ larger than the guide field,  $E_w (\om)  \geq B_0$, cannot propagate. {In extended systems, e.g.,  within magnetospheres of neutron stars, this leads to opening of the \ms\  by a strong wave.}
 \end{abstract}

\section{Introduction}
\label{introduction}
Dynamics of nonlinear waves in plasmas is a classical problem in plasma physics  \citep{AkhiezerPolovin,1975OISNP...1.....A,1974JPlPh..12..297C}.  Recently it became important for astrophysical Fast Radio Bursts \citep[FRBs][]{2007Sci...318..777L,2022A&ARv..30....2P,2019ARA&A..57..417C}. 
 {FRBs are millisecond-long bursts of radio emission coming  from $\sim$ half way across the Universe. At the peak the (isotropic-equivalent) luminosity (in radio)  exceeds  billions of Solar luminosity (in optical) \citep{2022A&ARv..30....2P}.  
 
 An important astrophysical hint comes from the
observations of correlated radio and X-ray bursts from ultra-magnetized \NSs\ \citep{2020Natur.587...54C,2020Natur.587...59B}. This 
established the  FRB-magnetar connection. If the emission originates in the \mss,  the laser non-linearity parameter in these settings can be as high as staggering
 \be
 a_0 = \frac{e E_w}{m_e c \om}\sim 10^9
 \ee
 where $E_w$ is the electric field in the wave and $\om$ is waves' frequency. Parameter regime $a_0 \geq 1$ defines relativistic nonlinearity of the wave.

In addition, the wave-plasma interaction may occur in the highly magnetized environment, in the regime when the guiding \Bf\ strongly affects particle dynamics and laser-plasma interaction. Guide fields can be as strong, or even exceed, the quantum \Bf.
 In addition, plasma parameters are changing as the  EM pulse  propagates away from the \NS\  \citep[see \eg][]{2014ApJ...785L..26L,2014MNRAS.442L...9L,2016MNRAS.462..941L,2021ApJ...922L...7B,2021ApJ...922..166L,2023ApJ...957..102G}

Astrophysical challenges  of super-strong waves in plasmas connect to modern laser experiments.
The interaction of intense laser beams with plasma  is critical for the success of high energy density (HED) experiments, inertial confinement fusion, and, eventually, inertial fusion energy. 
High-intensity colliding laser pulses also can lead to a new type of laboratory experiments by creating abundant electron-positron pair plasma \citep{2006RvMP...78..591M,2008PhRvL.101t0403B,2012PhRvL.108p5006R,2020PhPl...27e0601Z}.
This process, first discussed more than a decade ago, is becoming a reality.

In addition, presence of pair plasma there is a number of principal differences between astrophysical and laboratory set-ups. (i) there are subluminal branches in astrophysical  highly magnetized plasmas; waves are {\it relativistically}  nonlinear. Modern lasers can in principle achieve $a_0 \geq 1$, but usually only in a very limited region of space (extended wave trains typically have $a_0 \leq  0$); (iii) almost universally, in laboratory experiments waves are shone onto plasma from outside, while in astrophysical setting nonlinear waves can be generated by particles  already inside plasma. These  correspond to  two different  set-up: boundary condition problem for waves falling from outside,, and eigenmode problem  for wave inside plasma.

Finally, there is a number of limitations to numerical PIC approaches. High intensity EM waves with $a_0 \gg 1$ falling onto pair plasma accelerate it ponderomotively to relativistic energies,  and experiences modulational instability that leads to the reflection \citep[][these works explored unmagnetized plasmas]{2025arXiv250906230T,2025arXiv250920594L}.  Finally, setting-up a subluminal wave already inside plasma is highly nontrivial in PIC simulations, as all the codes (to the best of our knowledge) ignore the initial current created by the initial particles' velocities (initial velocity is not used for current computation until the first update is done). This is especially problematic for subluminal waves, where $\nabla \times B \sim (4\pi/c) {j} $. Additional numerical complication of setting nonlinear wave inside plasma involves the fact that velocity (current), \Bf\ and \Ef\ are all evaluated at different times. To set a correct self-consistent nonlinear wave inside plasma then would involve subtle half-step  corrections. These arguments demonstrate usefulness of purely theoretical approach.
}

In this paper, we address the basic properties of plasma in this new regime: dispersion relations of relativistically nonlinear waves. We consider a particular case of circularly polarized,  fully nonlinear EM wave propagating along \Bf. This regime allows, within the two-fluid approximation, a fully nonlinear treatment.

We employ a two-fluid (cold and collisionless) approach. An important advantage of the two-fluid treatment over the MHD approach \citep[\eg][]{2012A&A...542A.128H} is that the current is calculated self-consistently from the dynamic equations for each species.

\section{Two-fluid  model  for relativistically nonlinear  circularly polarized waves }

 {
The governing equations within the two-fluid  model  for relativistically nonlinear  circularly polarized (CP)  wave are exceptionally simple. They
 include just Maxwell's equations and transverse   force balance}
\ba &&
\nabla \times \B= {4\pi}\J+  \partial_t \E
\nn &&
{\bf J} = e  ({\bf v}_p -{ \bf v}_e) n
\nn &&
d_t {\bf p} _{e,p} = \mp \left( \E + \frac{{\bf p} _{e,p}}{\sqrt{1+p_{e,p}^2}}   \times \B \right) 
\label{aaa}
\ea
where $ {\bf p} _{e,p} $ are particles' momenta; upper signs are for electrons. {Equations are written in the common gyration frame (no axial motion).}

{
This major simplification comes from the  fact that  
for a fully  relativistically nonlinear  CP EM wave propagating along \Bf\ (in $z$ direction), the oscillations are harmonic (this is not true for { linearly polarized, LP }),  while density remains constant. This is a major observation underlying the present work.
}
 
 Let us introduce a unit vector corresponding to the waves' vector potential
\be 
{\bf e} _w= \left\{\cos \left( \om t- k z\right),-\sin \left(\om t- k z\right),0\right\}
\label{pi}
\ee

Motion of particles in the \EM\  fields (guide plus waves)  consists of forced motion by the wave and free oscillations.
Importantly, we are interested in forced oscillations due to the influence of the wave on particle motion. (Though this statement seems obvious, it's an important point in selecting the correct dispersion branches, see \S \ref{negative}.)
In this case, for CP, the momentum of the particle is parallel to the waves' \Bf\  (aligned or counter-aligned).  Thus we can write
\be
{\bf p} _{e,p} = a _{e,p}  {\bf e} _w
\ee

The  relations (\ref{aaa})   then give
\ba &&
n^2-1= \left(\frac{a_p}{\sqrt{1+ a_p^2}}-\frac{a_e}{\sqrt{1+ a_e^2}}\right) \frac{\omega _p^2 }{ a_0 \om^2} = \left(\tanh \chi _p-\tanh \chi _e\right)  \frac{\omega _p^2 }{a_0 \omega
   ^2}
\\  &&
{a_0}= \left( \frac{f_B}{\sqrt{1+ a_p^2}}-1 \right) a_p  =\tanh \chi _p \left(f_B-\cosh \chi _p\right) 
\\ &&
{a_0}= \left( \frac{f_B}{\sqrt{1+ a_e^2}}+1 \right) a_e =\tanh \chi _e \left(f_B+\cosh \chi _e\right)
\\ &&
a_{p,e} \equiv \sinh (\chi_{e,p} )
\label{disp}
\ea
where  $n=k/\om$,  frequency parameter
$
f_B = {\om_B}/{\om}
$,
 and we introduced rapidity $\chi_{e,p}$ (so that $v_{e,p} = \tanh (\chi_{e,p} )$). 

{
In Eq. (\ref{disp}) the plasma frequency
\be 
\om_p^2 = \frac{4 \pi e^2 n}{m_e}
\ee
 is defined with respect to the density of each component (\eg\ in pair plasma the relevant terms in the dispersion relations are $ \propto 2 \om_p^2$). The 
cyclotron frequency 
\be 
\om_B = \frac{e B}{m_e c}
\ee
is positively defined; the signs of charges are explicitly taken into account. 

In what follows, 
whenever dimensionless momentum or energy appear, they should be understood in terms of $p/(m_e c)$ and $\epsilon /(m_e c^2)$. The speed of light and elementary charge are set to unity. 
When we refer to the electron-ion case (e-i), the ions are assumed to be motionless ($m_i \to \infty $  limit). To keep the notations consistent,  electron quantities come with a subscript $e$, while ion and positron quantities come with a subscript $p$.
}

By our choice of polarization and direction of \Bf, quantity $a_e$ is always positive, while  $a_p$ can have both signs. This is related to the possibility of cyclotron resonance: by the choice of polarization, it is the positively charged particles (positrons) that can be in resonance. Below we refer to them as ``resonant particles" - this term is related to the type of particles, not  particular  
particles that are in resonance.

Given our choice of polarization, $|\chi_p| > \chi_e$. Thus, superluminal waves correspond  to $\chi_p < 0$, while  subluminal to $\chi_p > 0$.

 \section{Basic linear plasma waves propagating along \Bf}
\begin{figure}
\includegraphics[width=.99\linewidth]{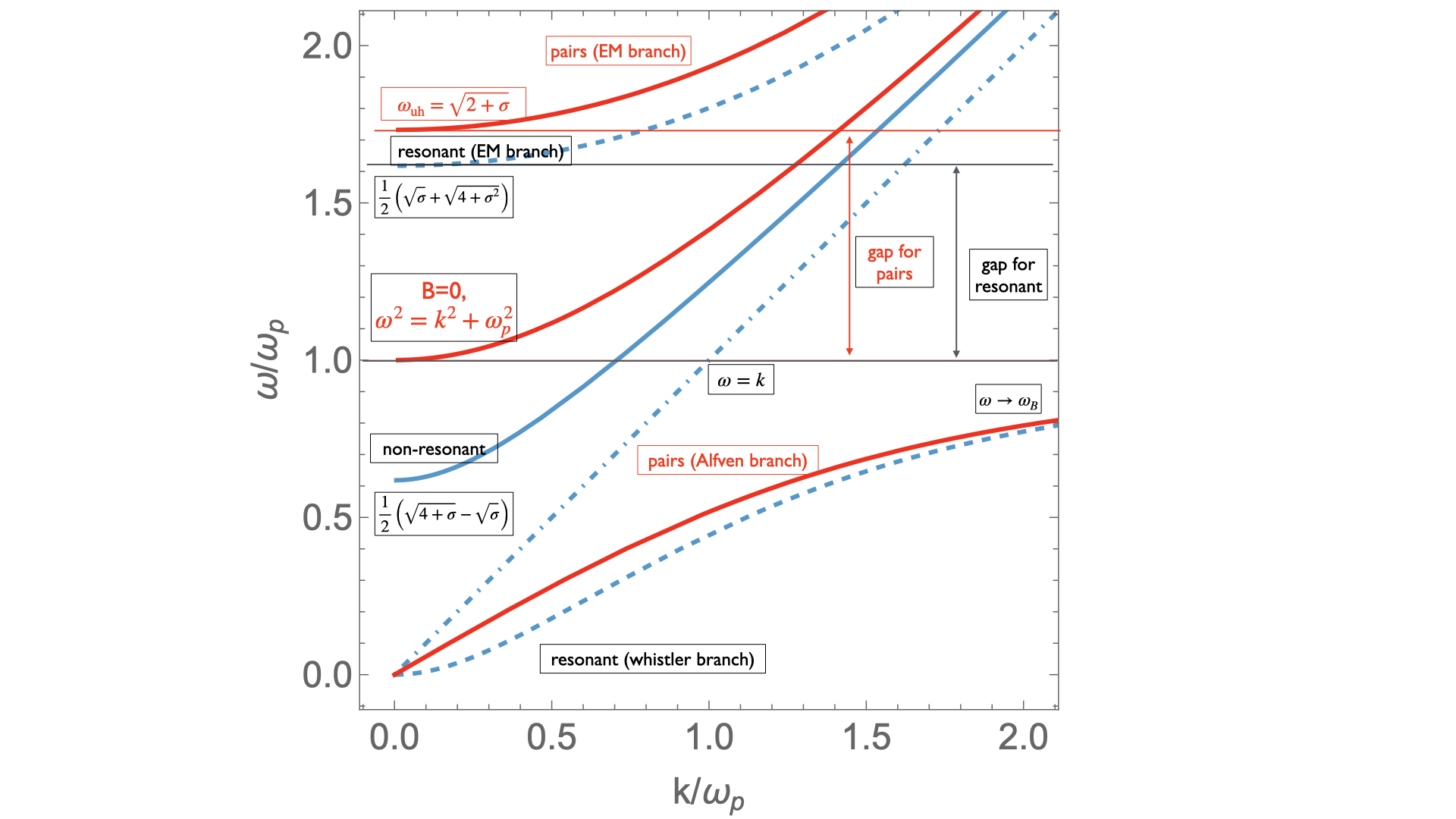}
\caption{ Dispersion relations for linear waves for particular choice $ \sigma \equiv  (\om_B/\om_p)^2=1$.  
 } 
\label{figlinear}
\end{figure}

As we are interested in relativistic modifications, let us briefly review the conventional linear case { \citep[\eg][]{1975OISNP...1.....A}}, see Fig. \ref{figlinear}, in which case fluctuating quantities and dispersion relation are
\ba &&
v_{0,e} = a_0 \frac{\om}{\om+\om_B}
\nn &&
v_{0,p} = a_0 \frac{\om}{\om_B- \om}
\nn &&
|v_{0,p}| > v_{0,e}
\nn &&
r_p = \frac{|v_{0,p}| }{\om} > r_e = \frac{|v_{0,e}| }{\om}
\nn && 
n^2 -1= - \left( \frac{1}{\omega \left(\omega _B-\omega \right)} + \frac{1}{\omega \left(\omega _B+\omega \right)}\right) \om_p^2 \overset{\rm pairs}{\to}  \frac{2 \om_p^2}{\om_B^2-\om^2 } 
\label{linear}
\ea
where $r_{e,p}$ are corresponding Larmor radii. In Eq. (\ref{linear}), we explicitly separated contributions from resonant and non-resonant particles. The final expression, after  $\to$  sign, is for a pair plasma. Notice that at $\om> \om_B$ velocities of particles are counter-aligned { with respect to each other}, so that their currents add, while for  $\om< \om_B$ velocities of particles are aligned so that their currents subtract.  
For definiteness, we label the direction of rotation with respect to non-resonant particles, so that  "co-rotating resonant particles" means the velocities of two species are aligned (in the linear case, this occurs for $\om_B > \om$, see Eq. \ref{linear}).

The subluminal branches, whistler and/or \Alfven, extend to $0<\om<\om_B$.
At small $\om\ll \om_B$,
\ba &
\om = \frac{k^2 \omega _B}{\omega _p^2}, & \mbox{whistlers}
\nn &
\frac{k^2}{\om^2} = 1+ \frac{2 \omega _p^2}{\omega _B^2} =  1+  \frac{2}{\sigma} , & \mbox{\Alfven}
\ea
where 
\be
\sigma =\frac{\om_B^2}{\om_p^2}
\label{sigma}
\ee 
is the  magnetization  parameter \citep{1984ApJ...283..694K}.

 Superluminal branches have a cut-off at
\ba &
\omega_{\rm uh}= \sqrt{\omega_B^2 + 2 \omega_p^2} = \sqrt{2+\sigma} \om_p,  & \mbox{pairs, upper hybrid frequency}
\nn & 
\om =\frac{1}{2} \left(\frac{\omega _B}{\omega _p}+\sqrt{4+ \frac{\omega _B^2}{\omega
   _p^2}}\right) =  \frac{1}{2}  ( \sqrt{4+\sigma} + \sqrt{\sigma} ) \om_p ,  & \mbox{resonant EM branch}
\nn &
\om= \frac{1}{2} \left(\sqrt{4+ \frac{\omega _B^2}{\omega
   _p^2}}- \frac{\omega _B}{\omega _p}\right) = \frac{1}{2}
    ( \sqrt{4+\sigma} - \sqrt{\sigma} ) \om_p,  & \mbox{non-resonant  EM branch}
    \label{cutofflin}
 \ea

\begin{figure}
\centering
\includegraphics[width=.49\textwidth]{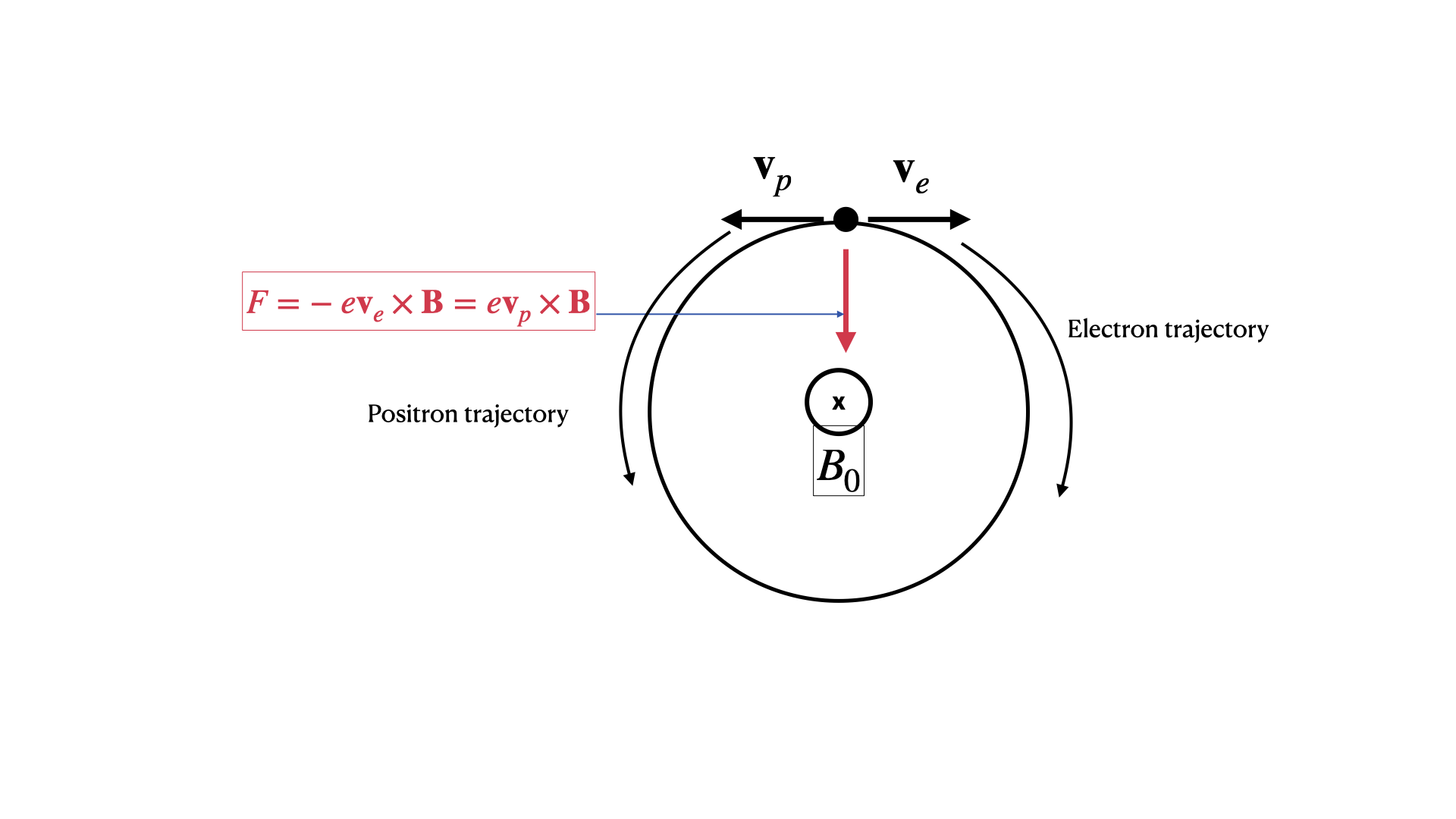} 
\includegraphics[width=.49\textwidth]{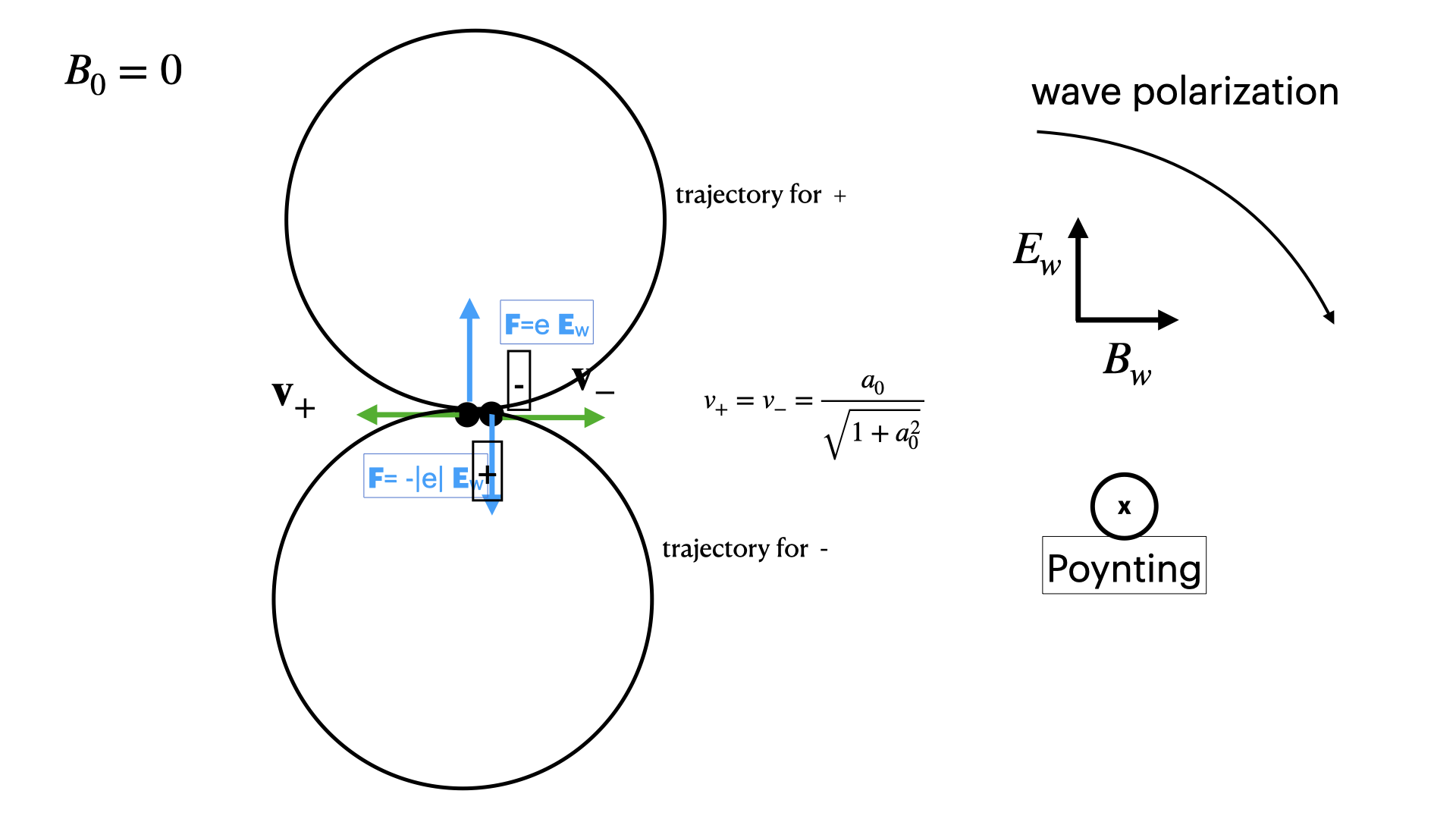} 
\includegraphics[width=.49\textwidth]{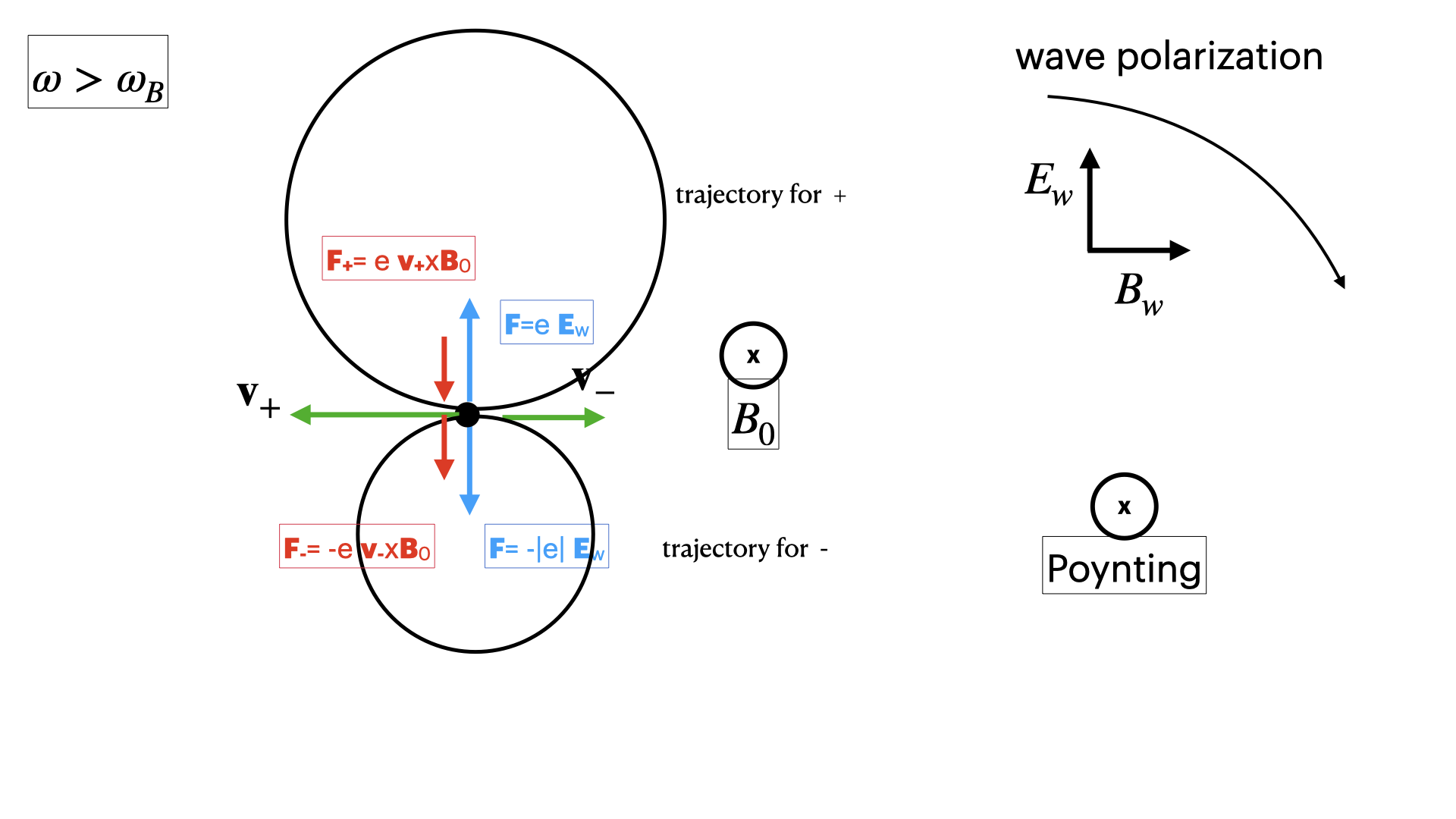} 
\includegraphics[width=.49\textwidth]{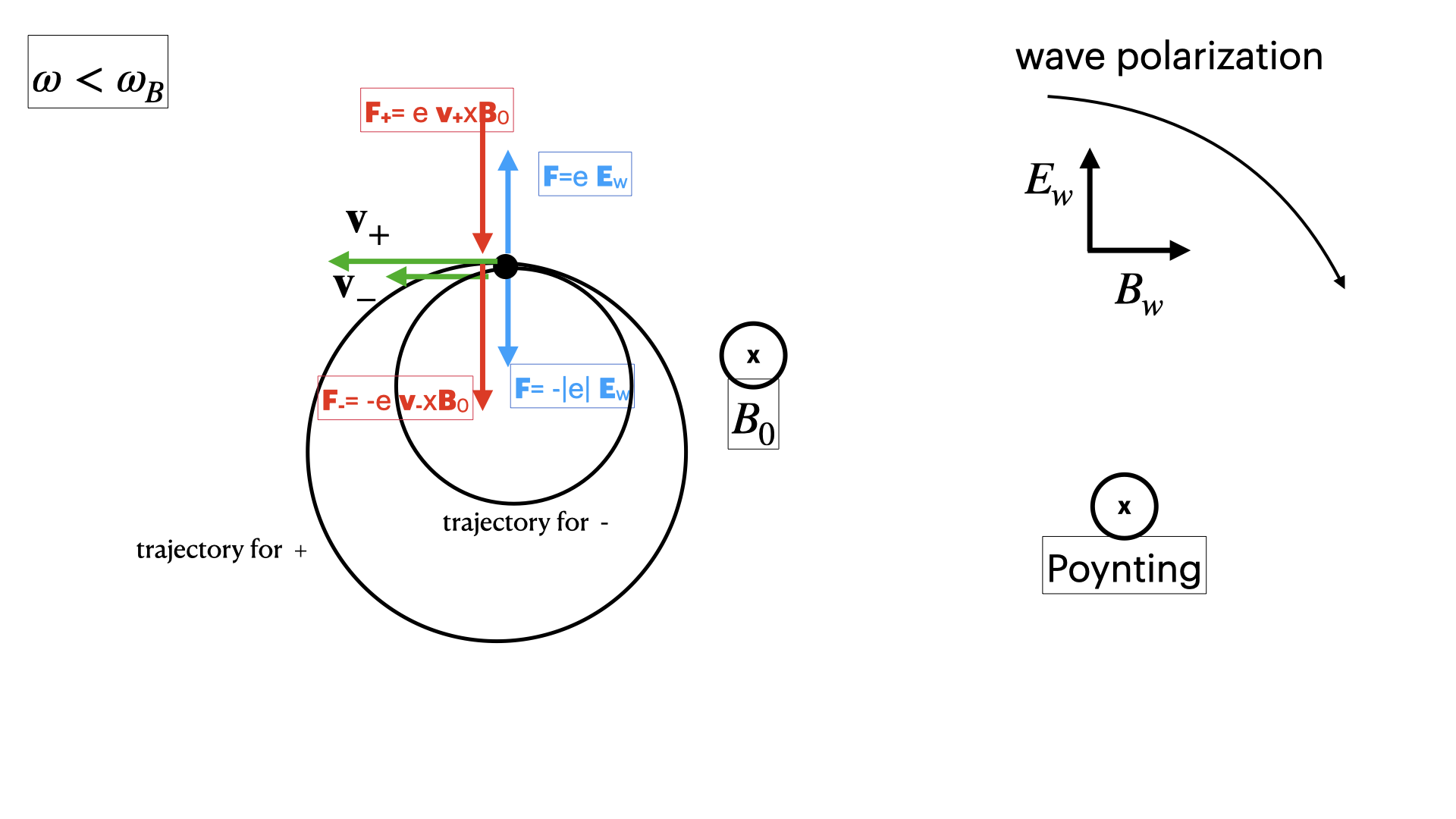} 
\caption{Particle trajectories in { circularly polarized  (CP)} EM waves propagating along the \Bf. Top left:  in external   \Bf\ without the wave; top right: in EM wave without guide \Bf.  In both cases at each point velocities of oppositely charged particles are counter-aligned (so the current add). Bottom row: trajectories in \Bf, left: weak \Bf\ (velocities counter-aligned, current add), strong \Bf\  (velocities aligned, current subtract). }
\label{traj} 
\end{figure}

Various particles' trajectories and forces are sketched in Fig. \ref{traj} (equal masses are assumed). There are two forces acting on a particle: \Ef\ of the wave $E_w$  and the Lorentz force from the guide field $v_{e,p} B_0$. For non-resonant particles, both forces add. For resonant particles they subtract: for $\om > \om_B$ the force from the electric field is larger, while for $\om < \om_B$ the Lorentz force dominates. 

\section{Single particle motion in non-linear CP EM wave with guide field}
\label{signgle}

For the CP wave, the dispersion relations and force balance equations separate. This allows fully nonlinear treatment. We first 
consider single particle dynamics in the presence of CP wave and guide \Bf\ (propagation along the field). For now, the frequency of the wave $\om$ is considered as given. In plasma, \S \ref{waves},  the wave frequency in the gyration frame has to be calculated self-consistently.

{
 A special type of solution (not a general one starting with arbitrary initial conditions)
involves particles' velocities instantaneous (counter)-aligned with the wave's \Bf.  The force  balance can be written in a fairly  compact form,
\be
\gamma_\pm m_e v _\pm \om= e ( E_w \pm  v _\pm B_0)
\label{Forces}
\ee
where $ \pm $ accounts for two directions of the background field/charge/polarization sign (speed of light is set to unity). But different realizations are fairly complicated depending on the values of the waves'  \Ef, guide $B_0$ and waves frequency $\om$. They are  illustrated  in  Fig. \ref{traj}.
}

  \begin{figure}
\includegraphics[width=.99\linewidth]{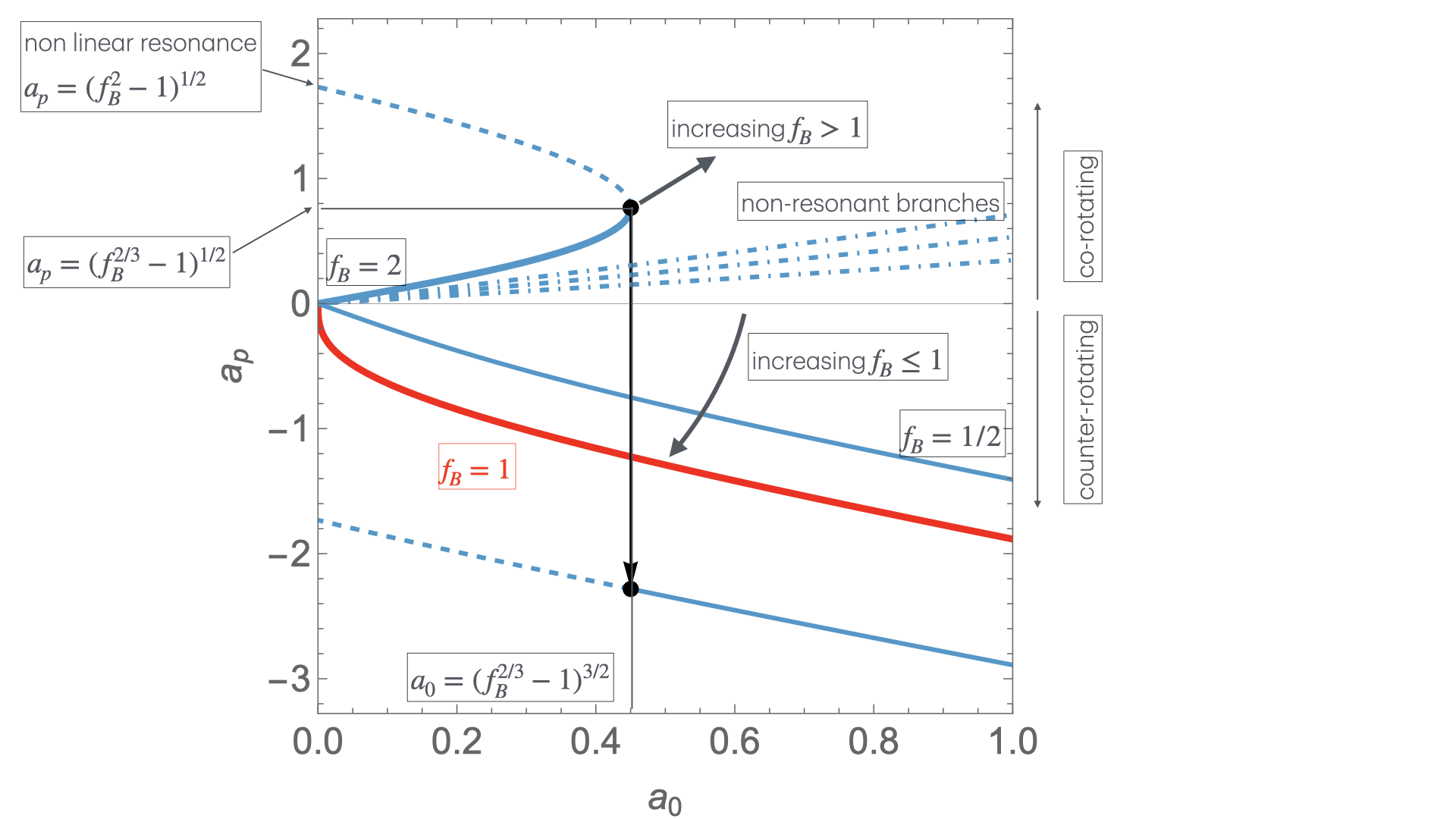}
\caption{ Particles' momenta in nonlinear CP wave. For $ \om \geq \om_B$ ($f_B= \om_B/\om \geq 1$, including the case $\om=\om_B$) resonant particles are counter-rotating (negative $\chi_p$). For   $ \om < \om_B$, resonant particles are co-rotating for mild $a_0\leq  (f_B^{2/3}-1)^{3/2}$ and counter-rotating for larger $a_0$. Dashed branches { are likely to be unstable}.
}
\label{singlepart}
\end{figure}

Momenta of non-resonant  $a_e$  (dot-dashed lines in Fig. \ref{singlepart}), are positive $a_e>0$  by our choice of polarization. Dependence of $a_e$ on parameters is fairly straightforward. 

The most interesting case is for particles that may be in resonance.   (Recall that in the linear case,  $ a_0 = a_p (f_B -1)$, the resonance is reached at $f_B=1$. Below the resonance  $a_p<0$ (counter-aligned with non-resonant particles), while above the resonance  $a_p>0$ (aligned).)

Nonlinear effects make the most important modifications for particle trajectories near the resonance $f_B \approx 1$. Formally, the resonance  shifts  from $f_B=1$  to 
\be
a_p = \sqrt{f_B^2-1} 
\label{gammap}
\ee
But, in fact, it is never reached, as we discuss next.

First, consider a special case $f_B=1$ (thick  line in Fig. \ref{singlepart}). In this case, momentum is determined from
\be
a_0=  \tanh (\chi_p )-\sinh (\chi _p)
\ee
This is an equation for $\chi_p(a_0)$. Since $a_0 >0$ by definition,  negative $\chi_p$ are needed (resonant particles are still counter-aligned with non-resonant, just like for $f_B < 1$). 

For small $a_0$,
\be
a_p = - (2 a_0)^{1/3}
\ee
Thus, for the special case $f_B=1$ particle motion {\it  is not}  a linear response to wave intensity in the small $a_0\ll 1$ limit.  For any $f_B \neq 1$,  the response is linear, $ |a_p| \propto a_0$ in the limit $a_0 \to 0$. 

The situation slightly above the resonance, $f_B \geq 1$, is more complicated:
\begin{itemize} 
\item At mild intensities, $a_0 \leq a_0^{(\rm crit)}= (f_B^{2/3} -1)^{3/2}$ there are three branches for the solution $a_p (a_0)$. The physical one (least energy of a particle and correct limit for $a_0 \to 0$) is {\it co-rotating} - positive $\chi_p$ (thick curve in Fig. \ref{singlepart}, near $f_B=2$ box). Thus, the synchrotron resonance (\ref{gammap})  is located on the upper curve, and hence is never reached. (We encounter here a mathematical oddity:  a formal solution for forced oscillations $a_p(a_0)$ connects to free oscillations $\om = \om_B/\gamma_p$ at $a_0 =0$).  {The non-linear cyclotron resonance (\ref{gammap}) corresponds to particles gyrating without a wave. Hence, this point does not reflect  a response of plasma to an \EM\ perturbation - it's an initial condition on particles' velocities. }
\item at intensities $a_0 \geq a_0^{(\rm crit)}$ there is only one counter-rotating branch (solid curve at bottom right in Fig. \ref{singlepart}. Thus, for fixed $f_B>1$ (below cyclotron resonance)  and increasing $a_0$ there is a transition (indicated by a vertical arrow) from a co-rotating to a counter-rotating state. Since counter-rotating states correspond to super-luminal waves,  for a given $a_0$ there are no sub-luminal waves beyond some $k$.
\end{itemize}

\section{Relativistically nonlinear  CP \EM\ waves in magnetized plasma}
\label{waves}

Next, we finally turn to the properties of nonlinear  EM waves in pair plasma. Above, in \S \ref{signgle}, we considered frequency as given. It is in fact, determined by the properties of plasma.

An additional complication comes from the arbitrary dependence of wave intensity on frequency: $a_0(\om)$ is a free parameter.
Generally, the dispersion relation has the form
\be
\om= \om ( k, a_0(\om))
\ee
 Various  $a_0(\om)$ will produce different dispersions $\om(k)$. 
 
In what follows, we first consider an exemplary case of  constant $a_0(\om)$. It highlights the key point: relativistically nonlinear subluminal modes terminate at some $\om^\ast-k^\ast$  point on the $\om-k$ plane. Later, in \S \ref{FixedEw}, we highlight general relations and discuss the case of constant ratio  $E_w/B_0$ (wave intensity to guide field).

The cases of single components and pair plasma are somewhat different. For the single-component case, the mathematics is simpler: eliminate particle momenta from the force balance, and use it in the expression for $n^2-1$. In pair plasma, the procedure is more complicated: particles' momenta $a_{e,p}$ both depend on the resulting waves' frequency. 

\subsection{Superluminal modes}

\begin{figure}
\includegraphics[width=.99\linewidth]{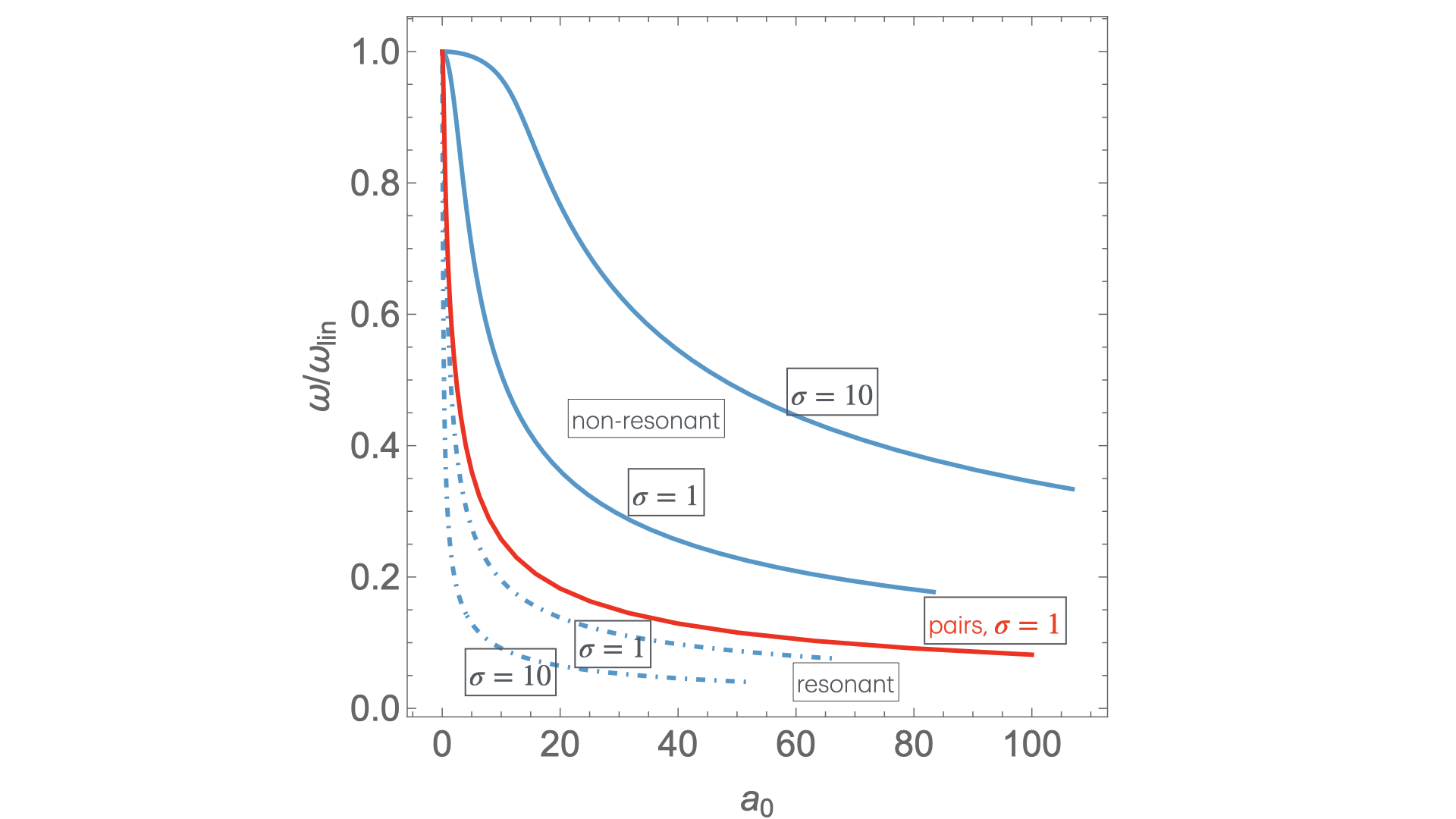}
\caption{ Nonlinear cut-off frequencies, as ratio to linear ones, Eq. (\ref{cutofflin}). In all cases, cut-off frequencies decrease with large  $a_0$. } 
\label{cutoffs-ao}
\end{figure}

Superluminal modes have $\chi_p <0$. 
All  superluminal modes (R, L and {\Alfven}) have cut-offs at 
\be
\om_{\rm cut} ^2 =  \frac{\tanh \chi _e-\tanh \chi _p}{a_0}  \om_p^2,
\ee
Fig. \ref{cutoffs-ao}.

For the single-component cases,  these relations can be resolved analytically. The resonant ($\chi_p <0$) case is:
\ba &&
\om_{\rm cut} ^2 =  \frac{\tanh|\chi _{p}|)}{a_0}  \om_p^2 = \frac{\omega _p^2}{\cosh \left(\chi _p\right)-f_B}=\frac{1}{2} \text{sech}\left(\chi _p\right) \left(\sqrt{\omega _B^2+4 \omega _p^2 \cosh
   \left(\chi _p\right)}+\omega _B\right)
\nn &&
a_0=f_B \tanh \left(\chi _p\right)-\sinh \left(\chi _p\right) = \sinh \left(\chi _p\right) \left(\frac{2 \omega _B}{\sqrt{\omega _B^2+4 \omega _p^2
   \cosh \left(\chi _p\right)}+\omega _B}-1\right)
\ea
and non-resonant:
\ba &&
\om_{\rm cut} ^2 =  \frac{\tanh\chi _{e}}{a_0}  \om_p^2 = \frac{\omega _p^2}{\cosh \left(\chi _ e\right)+f_B}=\frac{1}{2} \text{sech}\left(\chi _e\right) \left(\sqrt{\omega _B^2+4 \omega _p^2 \cosh
   \left(\chi _e\right)}-\omega _B\right)
\nn &&
a_0=f_B \tanh \left(\chi _e\right)+\sinh \left(\chi _e\right)= 
\nn &&
 \frac{\omega _B \tanh \left(\chi _e\right) \left(\sqrt{\omega _B^2+4 \omega _p^2
   \cosh \left(\chi _e\right)}+\omega _B\right)}{2 \omega _p^2}+\sinh \left(\chi
   _e\right)
\ea
\citep[compare with Eq. (8.1.4.2) in ][]{1975OISNP...1.....A}.

\begin{figure}
\includegraphics[width=.99\linewidth]{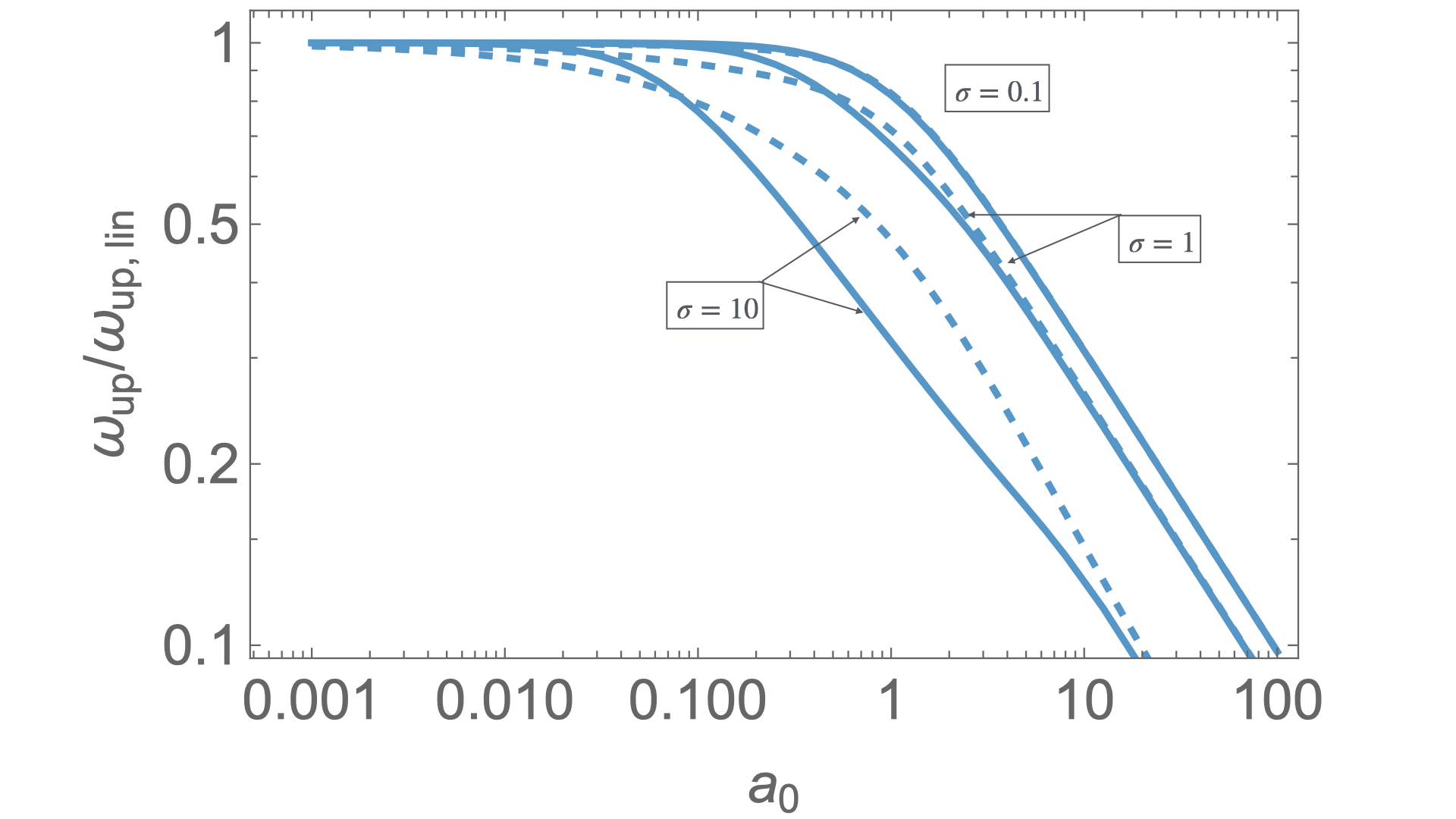}
\caption{ Nonlinear cut-off frequencies, as ratio to linear ones. Solid lines are solutions of  Eq. (\ref{cutofflin}), dashed lines are approximations  (\ref{approx}). In all cases, cut-off frequencies decrease with large  $a_0$. } 
\label{cut-off-compare}
\end{figure}

For pair plasma, the cut-off frequencies cannot be put into a compact expression. A good approximation is achieved if we use 
\ba &&
\om_B \to \om_B  /  ( 1+a_0^{2/3})^{3/2}
\nn &&
\om_p \to \om_p/(1+a_0^2)^{1/4}
\label{approx}
\ea
see Fig. \ref{cut-off-compare}. The relation for $\om_p$ is exact for $\om_B=0$ case \citep{AkhiezerPolovin}.

\begin{figure}
\includegraphics[width=.99\linewidth]{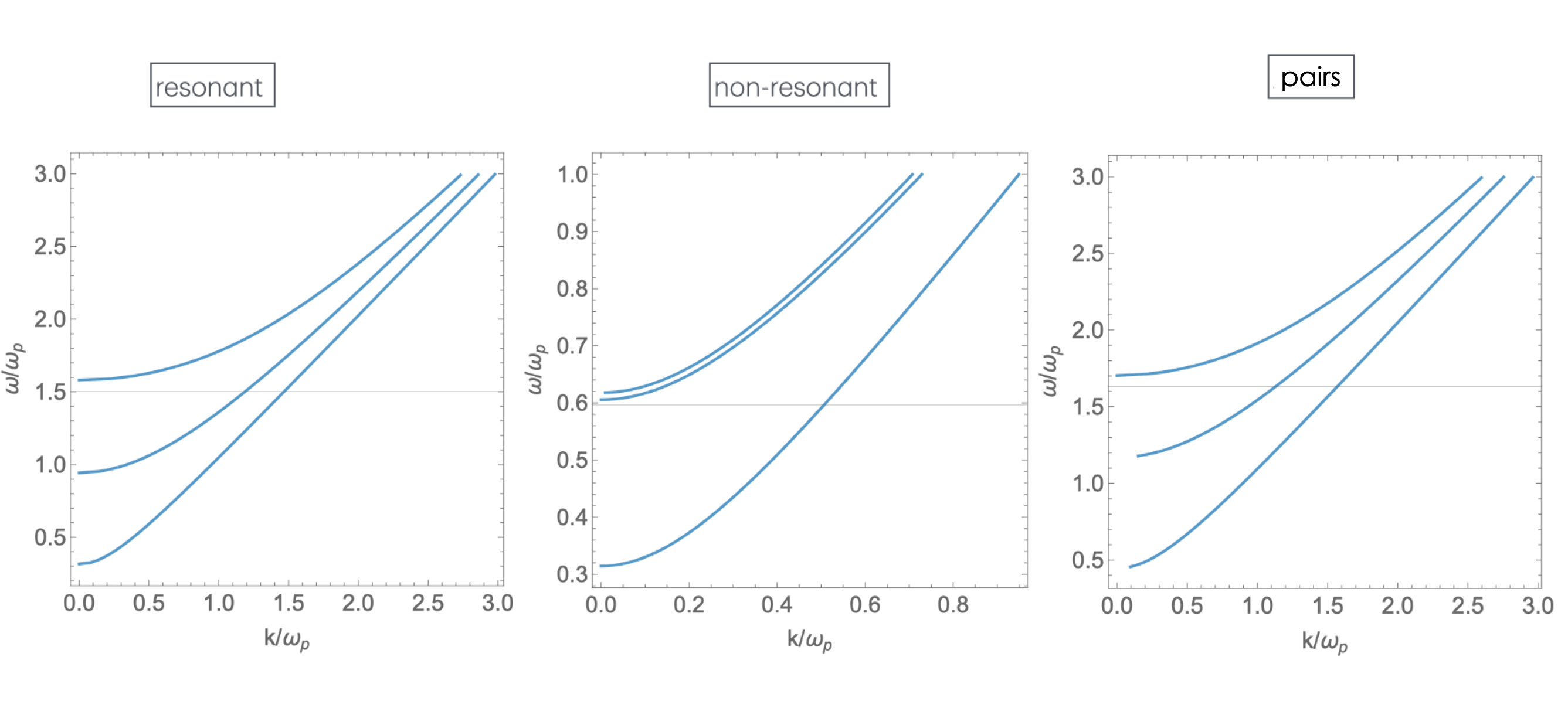}
\caption{Dispersion  curves for super-luminal  modes for  $\sigma=1$, $a_0 =0.1,\,1,\, 10$ (top to bottom).} 
\label{disp-super}
\end{figure}

In Fig.\ref{disp-super}, we plot dispersion curves for super-luminal modes. Qualitatively, the curves look similar to the linear case, with cut-off frequencies shifted down for large $a_0$.

\subsection{Subluminal modes}

 There are two types of subluminal modes:  whistlers (single-component, resonant), and \Alfven modes (pairs); there are no subluminal modes corresponding to single-component non-resonant particles.

\begin{figure}
\includegraphics[width=.49\linewidth]{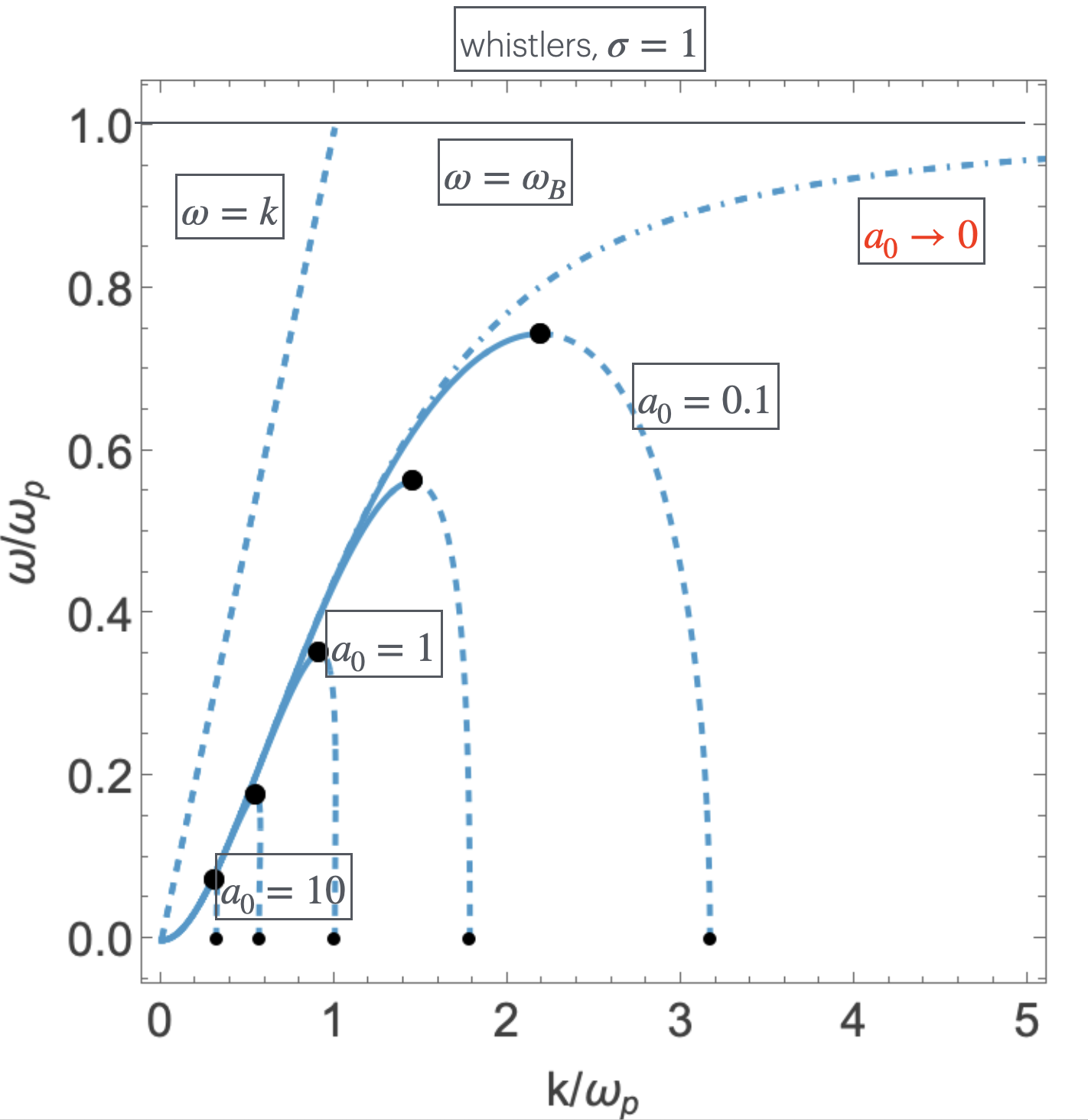}
\includegraphics[width=.49\linewidth]{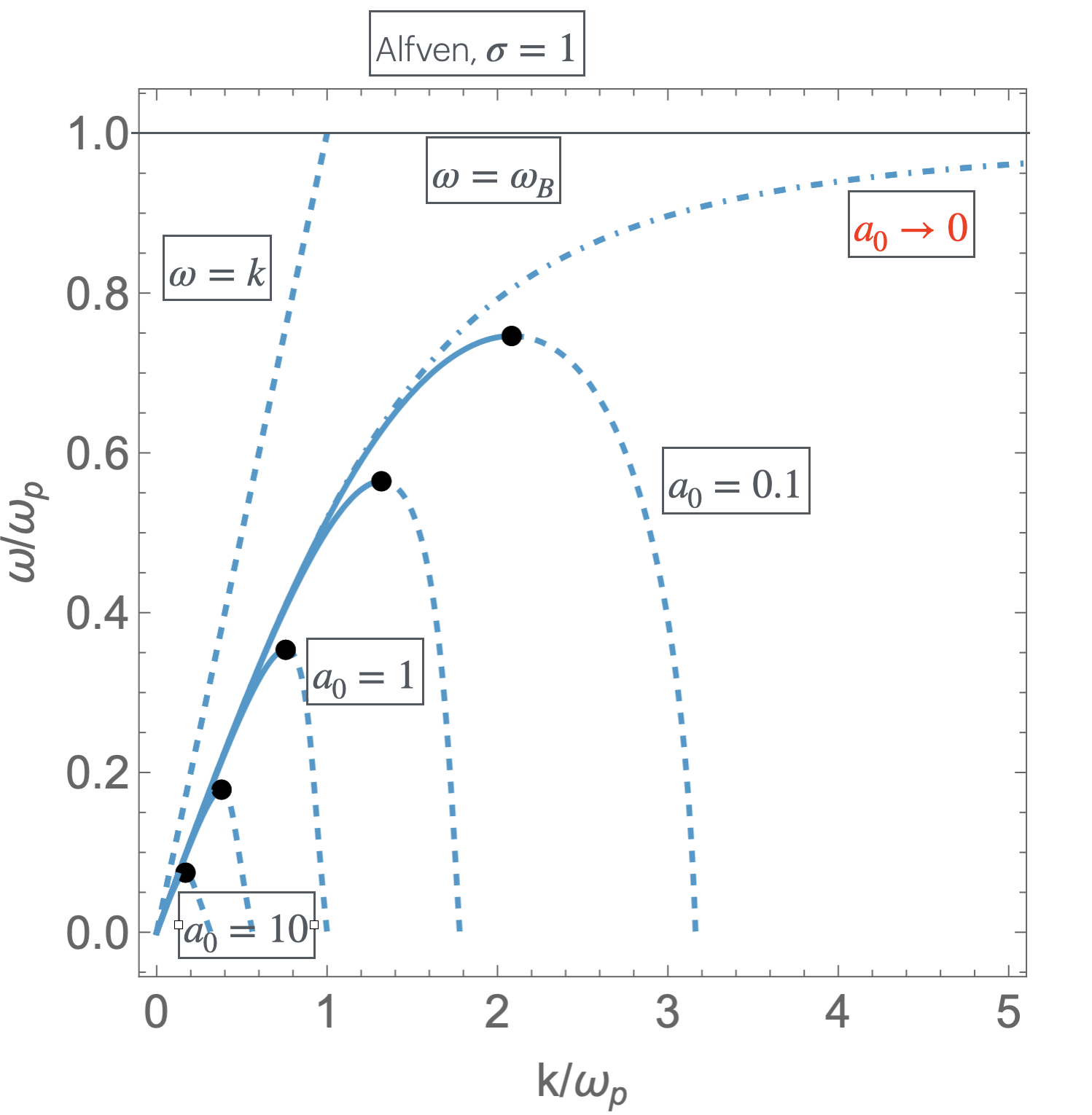}
\caption{ Dispersion curves for relativistically nonlinear whistlers (left panel) and \Alfven waves (right panel), $\sigma =1$.  Dispersion curves experience a bend at finite $k^\ast- \om^\ast$; above this point, resonant particles are on the upper branch in Fig. \ref{singlepart}.  { Dispersion curves terminate at $\om=0$,  $k_{\rm max} = \om_p/\sqrt{a_0}$ (low row of points)}.}
\label{sub}
\end{figure}

 {\it  A qualitatively new effect appears for subluminal modes - bend of the dispersion relation at finite $\om^\ast-k^\ast$, Fig. \ref{sub}.}  At the bend the group velocity  $v_g = \partial \om / \partial k$ becomes zero. Above the bend  (at higher $k$), the group velocity is negative.

  Recall that subluminal modes correspond to $f_B \geq 1$, while the resonant component has $\chi_p \geq 0$. With increasing $a_0$ it follows the middle branch in Fig. \ref{singlepart}. The  middle branch  $a_p (a_0)$ terminates at some finite $a_0$.  The critical point corresponds to 
 \be 
 \om^\ast= \frac{\omega _B}{\left(1+ a_0^{2/3}\right){}^{3/2}} \to  \frac{\omega _B}{a_0}
 \label{omegastar}
 \ee
(the latter relation is for $a_0 \gg 1$).  It is the same for both whistlers and \Alfven modes.

For fixed $a_0(\omega)= $ const,  
dispersion curves bend over in the $\om-k$ plane at some finite values of $\om^\ast-k^\ast$, and since above the bend modes are likely to be unstable, the dispersion curve effectively terminates. At frequencies well  below the cyclotron frequency,  $f_B \gg 1$, Alfven waves exist only for $\om \leq \omega _B/ a_0$, or reverting to the fields,
\be
E_w \leq B_0
\label{Ew} 
\ee
(see \S \ref{FixedEw}).

 For single-component plasma (relativistically  nonlinear whistlers), the corresponding wave vector and phase velocities  are 
 \ba &&
 k^2 = \omega ^2-\frac{\omega  \omega _p^2}{\omega  \sqrt{1+ a_p^2}-\omega _B}
 \nn &&
 k^{\ast, 2} =
 \frac{\omega _B^2}{\left(1+a_0^{2/3}\right){}^3}+\frac{\omega
   _p^2}{\sqrt{1+a_0^{2/3}} a_0^{2/3}}
   \nn &&
   v ^{\ast, 2} = \frac{a_0^{2/3} \sigma }{a_0^{2/3} \sigma +\left(1+ a_0^{2/3}\right){}^{5/2}} \to \frac{\sigma}{a_0 + \sigma}
   \ea
   ($ v ^\ast$ is the terminal phase velocity of whistlers),  Fig. \ref{sub} left panel.
   
 For relativistic whistlers, the  maximal phase velocity 
   is reached at 
   \ba &&
   a_0 = (2/3)^{3/2} = 0.544
   \nn &&
    v ^{\ast}_{\rm max} = \sqrt{\frac{\sigma }{\sigma +{25 \sqrt{\frac{5}{3}}}/{6}}}
    \nn &&
    \gamma^{\ast, 2} =1+ \frac{6}{25} \sqrt{\frac{3}{5}} \sigma
    \ea
 
{  The dispersion curve ends at $\om=0$, $k_{max}=\om_p/\sqrt{\alpha_0}$ (low row of points in Fig.  \ref{sub}). }

    In pair plasma $ k^{\ast}$, terminal wave vector for relativistically nonlinear \Alfven waves) has to be found numerically, Fig. \ref{sub} right panel.

\begin{figure}
\includegraphics[width=.99\linewidth]{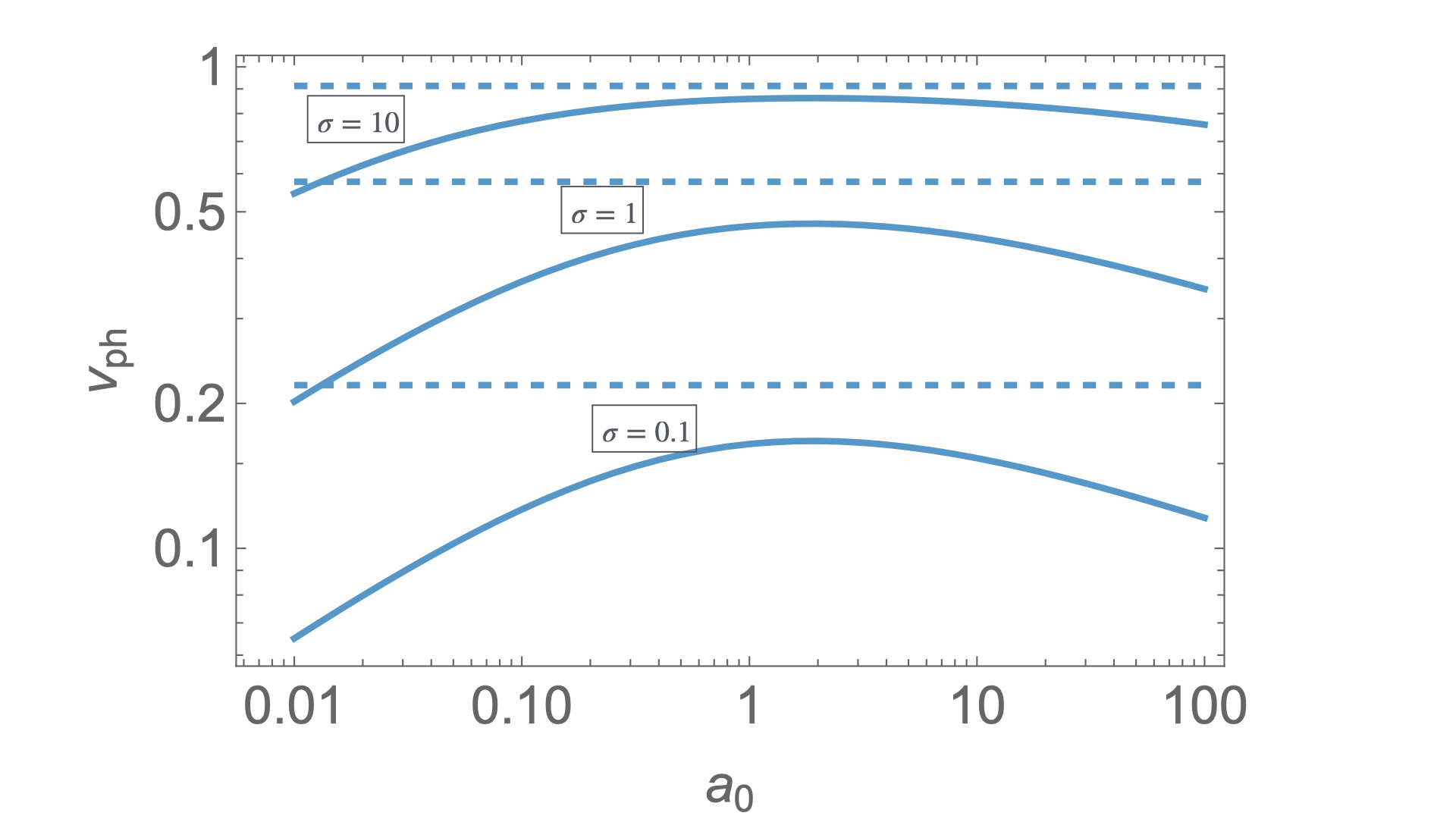}
\caption{Terminal phase velocities for \Alfven waves, $\sigma=1$. Dashed lines are linear limits for $k\to 0$,  $v_A = \sqrt{\frac{\sigma }{2+ \sigma }} $. } 
\label{terminalAlfven}
\end{figure}

In Fig. \ref{terminalAlfven} we plot terminal \Alfven velocities as a function of $a_0$ for different magnetizations. For comparison, in dashed lines we should linear \Alfven speed for $\om\to0$, 
$v_A = \sqrt{\frac{\sigma }{2+ \sigma }} $ (factor $2$ accounts for two species, $\sigma$ is defined with respect to each separately).  Relativistically nonlinear \Alfven waves are slower due to increasing effective mass.

\subsection{Negative group velocity branch and the bend in $\om(k)$}
\label{negative}

Negative group velocity branches, at $k> k^\ast$,  are likely to be unstable. The stability criterion for {\it linear} waves \cite[][parags. 80 and 84 ]{landau_60} is violated
\be
\partial_\om  (\om \epsilon (\om)) = -  n^2 - 2 n (\partial_k \om)^{-1} <0
\label{stab}
\ee
This indicates that the wave has negative energy and is likely to become unstable by coupling to the positive energy modes. Negative energy of the wave indicates that the wave would take energy from the plasma components.

As Fig. \ref{singlepart} indicates, at any given $a_0$, the upper branch has particles with larger momenta than the lower branch. It's an interesting coincidence that dispersion curves, which describe the response of a medium to a perturbation, terminate at free-rotating particles. 

{
As another argument in favor of  instability of a branch above $k> k^\ast$, the terminal  point $\om=0, \, k \neq 0$ corresponds to the conditions for Electron Cyclotron Maser/gyrotron instability \citep{Melrosebook}, with the exception that  in pair plasma both components have distribution in perpendicular momenta $\propto \delta ( a_{e/p} - p_{0, e/p})$. This clearly indicates population inversion and instability. 
Finally,   negative group velocity is possible in media with inverted populations, and hence unstable \citep{1970PhRvA...1..305G}. 
}

In passing, we note that there are in fact media where the phase and group velocities of electromagnetic waves are oppositely directed: they go under the name left-handed media or negative-index metamaterials \citep{Veselago_1968}. The difference is that experimental setups with $n<0$ are fixed, while here they are dynamic.
Another prominent case when phase and group velocities are counter-aligned is the Cherenkov emission \cite[\eg][]{Bolotovskii&Stolyarov}. 

\subsection{Dispersion relations for   arbitrary    $\eta_w = E_w/B_0$}
\label{FixedEw}

Above, we considered a case of constant $a_0(\om) \equiv a_0$, and hence increasing waves' \Ef\  $E_w(\om) \propto \om $. We found that  \Alfven waves dispersion terminates at $E_w \sim B_0$.  Next, we discuss generation properties of the dispersion relation, and repeat calculations for  arbitrary  $\eta_w = E_w/B_0$. 

{
Instead of (\ref{disp}), for given  $\eta_w$ we find
\ba &&
n^2-1= \left(\frac{a_p}{\sqrt{1+ a_p^2}}-\frac{a_e}{\sqrt{1+ a_e^2}}\right) \frac{1}{\eta_w}  \frac{\omega _p^2}{\omega   \omega _B} = \left(\tanh \chi _p-\tanh \chi _e\right) \frac{1}{\eta_w}  \frac{\omega _p^2}{\omega   \omega _B} 
   \nn && 
\eta_w= \left(\frac{1}{\sqrt{1+a_p^2}}-\frac{\omega }{\omega _B}\right) a_p  =\tanh \chi _p  - \frac{\sinh \chi _p}{f_B}
   \nn &&
\eta_w= \left(\frac{1}{\sqrt{1+a_e^2}}+\frac{\omega }{\omega _B}\right)  a_e =\tanh \chi _e  + \frac{\sinh \chi _e}{f_B}
\label{disp1}
\ea
}

For frequency-dependent $\eta_w(\om)$, the relation (\ref{omegastar}) for the critical frequency, where dispersion of subluminal modes terminates,  can be written as 
\ba &&
\tilde{\om}^{\ast} = \frac { \om^\ast}{ \om_B}=  \left(1 - \left( \eta_w  \right) ^{2/3}  \right) ^{3/2}
 \nn &&
 a_p^\ast= \sqrt{\frac{1}{ \tilde{\om}^{\ast, 2/3}}-1} = \frac{  \eta_w^{1/3}}{\sqrt{1-\eta_w^2}}
 \nn &&
 \eta_w (\om) = \frac{ E_w(\om) }{B_0}
 \label{omegastar1}
 \ea

\begin{figure}
\includegraphics[width=.49\linewidth]{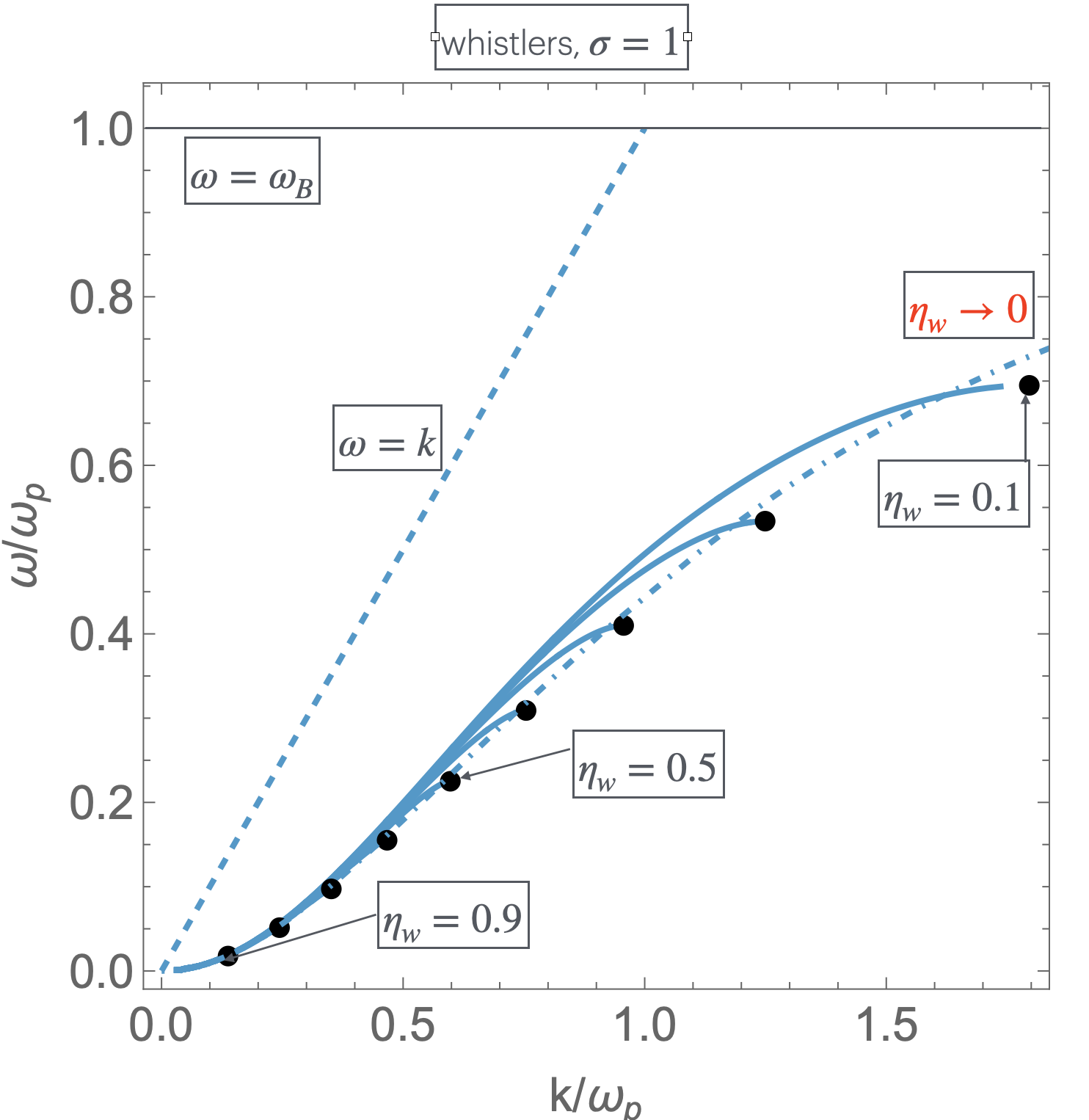}
\includegraphics[width=.49\linewidth]{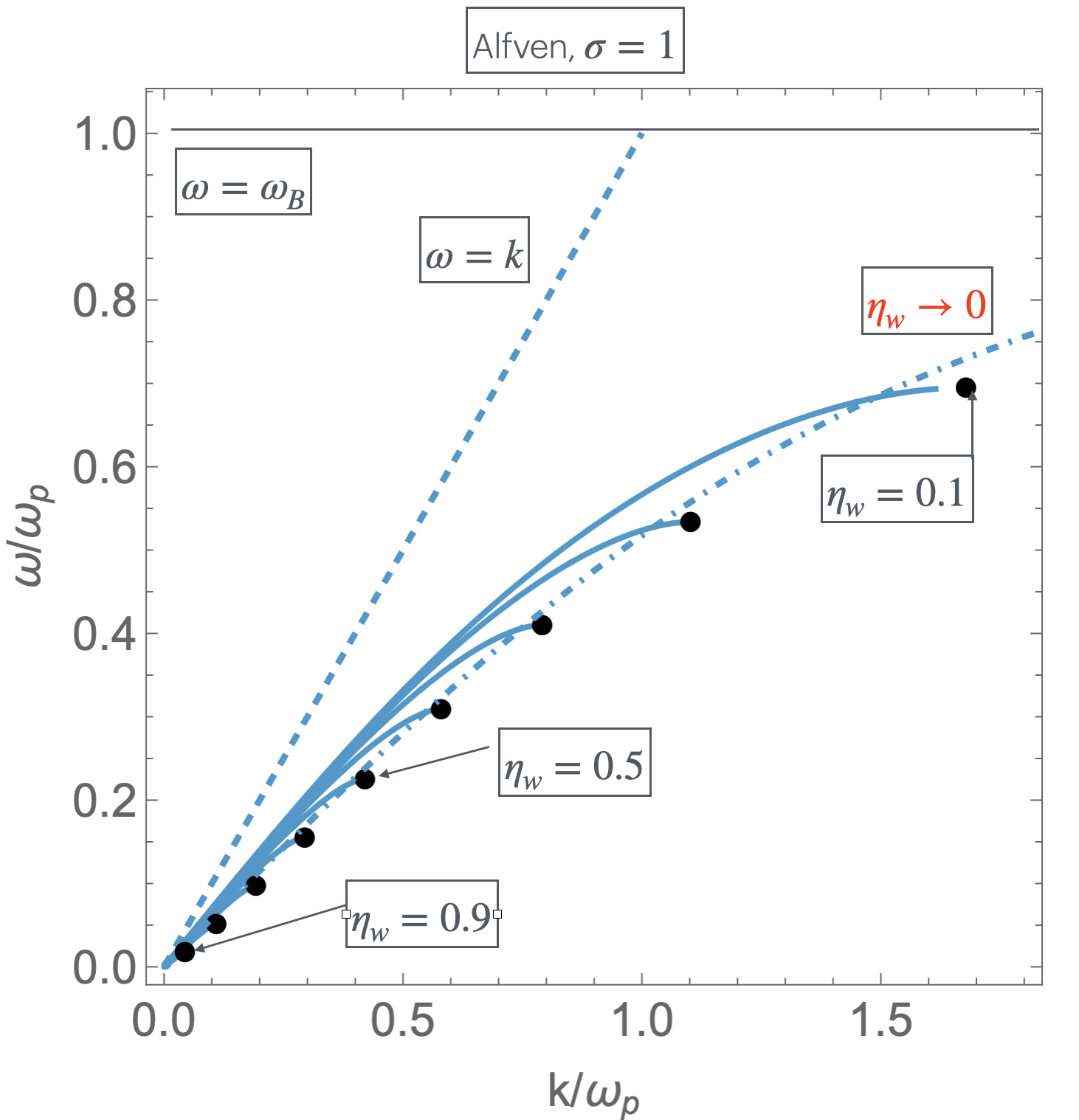}
\caption{Same as Fig. \ref{sub}, assuming  constant ratio of wave intensity to guide field $\eta_w = E_w/B_0$.}
\label{constantEw}
\end{figure}

 {
 In Fig. \ref{constantEw} we plot the dispersion curves for relativistically nonlinear subluminal waves assuming a constant ratio of wave intensity to guide field $\eta_w = E_w/B_0$.  With increasing $\eta_w$,   propagating waves are confined to smaller corner near $\om,\, k \approx 0$. For $\eta_w \geq 1$ waves cannot propagate. 
 
Relation (\ref{omegastar1})  highlights the universal relationship: when at a given frequency the intensity of the fluctuating field in a subluminal wave exceeds the guide field, $\eta_w \to 1$,  the wave cannot propagate.

 }
 
\subsection{Finite ion mass}

{
In case of finite mass of the ion component, $\mu = m_p/m_e \neq 1, \infty$, general relations are much more complicated, as two new frequencies appear (ion cyclotron and  ion plasma), both modified by the nonlinear effects. 
To achieve relativistic motion of ions in unmagnetized plasma, a wave should be $ a_0 \sim  \mu  = 1836 $   ($a_0 $ is defined with respect to the electron mass). 

Introduction of guide field leads to a number of complications. Here the form (\ref{disp1})  provides a better insight.  First, equation for plasma response involves $\om_p^2/\om_B$ and is thus mass-independent. Mass dependence enters through dynamical equations for velocities of the two components.

As we discussed above, 
the most important point is the  fairly subtle  near-cyclotron-resonance behavior. 
Two cases can then be identified: CP waves that can resonate with electrons (typically called R-modes), and   waves that can resonate with ions (L-modes). 
R-modes are well described by the infinite mass limit considered above.

For L-mode, relation (\ref{omegastar1}) for the critical frequency, where dispersion of subluminal modes terminates,  can be written as 
\ba &&
\tilde{\om}^{\ast} _i= \frac { \om^\ast}{ \om_{B,i}}=  \frac { \om^\ast}{ \om_{B}} \mu = \left(1 - \left( \eta_w  \right) ^{2/3}  \right) ^{3/2}
 \nn &&
 a_p^\ast= \sqrt{\frac{1}{ \tilde{\om}_i^{\ast, 2/3}}-1} = \frac{  \eta_w^{1/3}}{\sqrt{1-\eta_w^2}}
 \nn &&
\om_{B,i} = \frac{\om_B}{\mu}
 \label{omegastar2}
 \ea
(the physical momentum of ions now is $\mu a_p$).
 
 An admixture of ions can have disproportionally large effect, as it introduces an  additional  low-frequency resonance. But calculating the properties of the dispersion relation would require a separate "tour de force" calculations.

}

\section{Discussion}

In this paper, we considered relativistically nonlinear circularly polarized waves propagating along \Bf. We were able to solve the system exactly, nonlinearly relativistic. These exact relations provide guidance to more general setups.

 Dispersion curves for single-component plasmas (two possible polarizations) and pair plasmas were investigated. For superluminal modes, the modifications from the linear case are qualitative: decreasing cut-off frequencies. 
 
{
The most interesting effect appears for subluminal modes: dispersion curves effectively terminate at some finite values of $\om^\ast-k^\ast$. At which point the group velocity becomes zero. At this points, the fluctuating \Ef\ of the wave becomes equal the guide \Bf. }

\begin{figure}
\includegraphics[width=.99\linewidth]{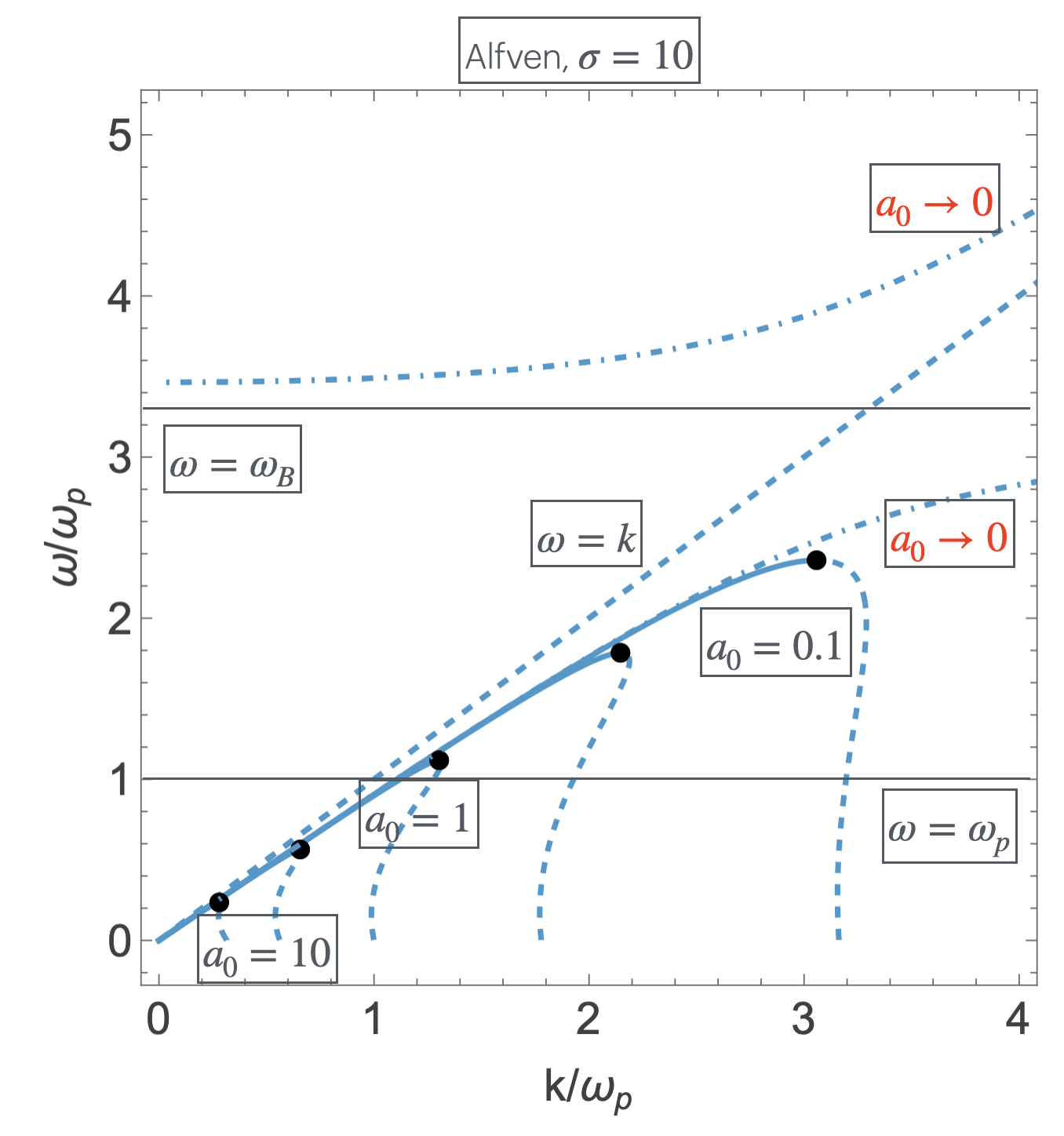}
\caption{Dispersion curves for $\sigma =10$. Compare with Fig. \ref{sub}. This illustrates that the critical  point is independent of the density. }
\label{Alfvensigma10}
\end{figure}

Though the effects reminds of charge starvation -  it is not.  The critical point is independent of plasma density, as illustrated by Fig. \ref{Alfvensigma10}.

The effect we observe is a variant of zero group velocity (ZGV). 
Waves at the ZGV frequency are stationary and do not propagate energy over a long distance; instead, the energy remains localized near the source, forming a standing wave. 



In astrophysical applications, condition  (\ref{Ew}) (or equivalently $\eta_w =1$, Eq.  (\ref{omegastar1})) may be reached for waves propagating in the dipolar field of magnetars' \mss.  At smaller radii, $B_0 \gg  E_w$, but since  $B_0  \propto r^{-3}$ while  $E_w \propto r^{-1}$,  condition  (\ref{Ew})   then  can be reached within the \ms.  { Instead of propagation, the waves will  pile-up near the critical point. As a result the wave will ``open'' the \ms\ \citep{2023MNRAS.524.6024S}.   The energy of the wave will be spent on distorting the \ms, which will then  recover on long resistive time scale. }

In pair plasma,  in  the  linear regime, there is a gap in dispersion relations for   $\om_B^2 \leq  \om^2  \leq  {\om_B^2 + 2 \om_p^2}$. In the relativistic regime,   there is still a gap
\be
\frac{\om_B^2 }{    ( 1+a_0^{2/3})^{3} } \leq \om^2 \leq \frac{\om_B^2 }{    ( 1+a_0^{2/3})^{3} }  + 2 \frac{\om_p^2 }{    \sqrt{  1+a_0^2}}
\ee
(upper limit is approximate). The width of the gap becomes smaller at large $a_0$,
\be
\Delta \om \sim \sqrt{2}  \frac{\om_p}{\sqrt{a_0}}
\ee

We stress that the solutions for nonlinear waves discussed here are applicable only to CP waves propagating along the \Bf.  In addition, we assumed that both species share the same gyration frame: this implicitly neglects possible effects of ponderomotive acceleration. (Electrons and positrons will experience {\it different} ponderomotive acceleration in a CP wave.)

{Finally, we hypothesize that waves will become modulationally  unstable before reaching the terminal point $\om^\ast - k^\ast$. This is better studies with direct PIC simulation (see \S \ref{introduction} for discussion of numerical challenges.)

\section{Acknowledgements}

 This research was supported in part by grant NSF PHY-2309135 to the Kavli Institute for Theoretical Physics (KITP). 
I would like to thank participants at the program "Frontiers of Relativistic Plasma Physics" for numerous discussions. Special acknowledgments are due to Tom Blackburn, Thomas Grismayer,  Pavel Kovtun, Yury Lyubarsky,  Mikhail Medvedev,   Luis Silva,  Anatoly Spitkovsky, Chris Thompson and Maria Vranic. I would also like to thank Sergey Bulanov  and Sergey Komissarov for comments on the manuscript.
\bibliographystyle{jpp}
 \bibliography{/Users/lyutikov/Library/CloudStorage/Dropbox/Research/BibTex,/Users/lyutikov/Library/CloudStorage/Dropbox/Research/BibTexShort.bib,//Users/lyutikov/Library/CloudStorage/Dropbox/Research/NASA_FRB.bib}

\end{document}